# A unified sedenion model for the origins of three generations of charged and neutral leptons, flavor mixing, mass oscillations and small masses of neutrinos

Qiang Tang[1*] and Jau Tang[2,3*]

[1]School of artificial intelligence, Anhui University of Science and Technology, Huainan, Anhui 232000, China

[2]Huazhong University of Science and Technology, Wuhan 430074, China

[3]Wuhan University, Wuhan 430072, China

Corresponding authors: Qiang Tang: tangqiang1992911@qq.com,     Jau Tang: jautang@hust.edu.cn

**Abstract**

We present a unified sedenion algebra model, without the need for an ad hoc Standard Model's hypothesis, we explain why there are three generations of charged and neutral leptons, why neutrinos have a vanishingly small mass, and why flavor-mixing and mass oscillations occur. We show that the sedenion algebra contains three types of non-associative octonion algebra, with each corresponding to a generation of leptons. By incorporating extra degrees of freedom, the generalized higher dimensional Dirac equation accounts for the internal structural dynamics. This study sheds light on the intrinsic physical properties of three generations of charged leptons and neutrinos, and the distinctive spacetime structures.

## 1. Introduction

The Standard Model [1] has been the cornerstone of subatomic particle physics for the last half-century. This model has successfully explained most physical properties of elementary particles and their composite particles with high accuracy. Despite its great success, many mysteries and unsolved problems remain. For example, it is unclear why there are three generations [2] of leptons, including the electron and its two heavier cousins, the muon and the tau. There exist also three generations of corresponding neutrinos with masses several orders of magnitude smaller [3]. In addition, there are experimental evidence of flavor mixing among these three types of neutrinos, and their mass-oscillation behavior [4] appears to contradict the Standard Model. Other than six types of leptons, there are also six types of quarks, up/down, charm/strange, and top/bottom quarks. The masses of each quark or lepton generation appear to follow Koide's simple empirical mass-ratio formulae [5, 6]. The masses of three generations of leptons, heavy quarks, and light quarks present an unsolved puzzle. These simple mass-ratio relations suggest that leptons and quarks are likely not point-like particles with no physical volume but have internal structural dynamics. There are many research directions aimed at solving these mysteries, including Clifford geometric algebra [7] with quaternions and octonions [8, 9], lattice gauge theories [10], grand unification theories [11], string theories [12], loops quantum gravity [13]. In this work, we shall focus on some of these puzzles and address only the unsolved issues related to the six leptons in the Standard Model.

## 2. Theory

This work presents a unified theoretical model of leptons that explains the existence of three generations of massive leptons and their corresponding almost massless neutrinos, as well as the flavor-mixing mechanism and mass-oscillation behavior. In the Standard Model, all six leptons are point-like objects with no physical volume, and they are described by the same Dirac equation [14] involving five Dirac gamma matrices. QED [15], built upon the Dirac equation, has been the most successful quantum field theory to accurately explain the electromagnetism of relativistic charged particles. This conventional QED, however, can neither explain why the mass of an electron must be about 0.511 MeV, nor why the fine structure constant should be ~ 1/137. The QED description for leptons in the Standard Model, however, cannot explain the origin of the three generations or the flavor mixing and mass oscillations[4] of neutrinos.

To overcome some shortcomings of the Standard Model and Dirac equation that assume a point-like particle, we propose a unified model, based on the geometric algebra operator approach. We use an octonion operator approach [16-18] that inherently includes three extra degrees of freedom to describe a particle's internal structural dynamics with respect to the reference frame for the particle's center of mass. Unlike the multiplication of two real numbers, complex numbers, matrices, or quaternions [16], the multiplication of octonion operators is generally not associative, except for the unit operator. A cleaver method to handle such non-associativity of octonions is using Zorn's construction scheme and multiplication rules [19]. We generalize the technique of Zorn's vector-matrix representation to operators composed of time-like and space-like tensor products of 4x4 matrices that are equivalent to bi-vectors in Gilford's geometric algebra formalism. Using this set of generalized matrix representations and multiplication rules, we can construct a coherent and unified theoretical model to describe all six types of leptons and their physical properties.

In Section 2, we will present a comprehensive theoretical framework for describing spin-1/2 particles. First, in Section 2.1, we will consider massless spin-1/2 particles using the operator approach. Next, in Section 2.2, we will focus on spin-1/2 particles with a rest mass. Then, in Section 2.3, we will introduce geometric algebra operators and provide an overview of octonion algebra and Zorn's vector-matrix representation of octonion operators, which satisfy the octonion multiplication rules with anti-commutative and non-associative properties. We will use these operators to construct a generalized Dirac equation incorporating three additional degrees of freedom to describe a particle's internal structural dynamics. In Section 2.4, we will present a more general sedenion model that accommodates three generations of massive leptons, namely, the electron, muon, and tau. Finally, in Section 2.5, we will present a model for three generations of neutrinos, and explain why their masses are almost zero, and the emergence of flavor mixing and mass oscillations that are beyond the Standard Model's description.

## 2.1. Massless spin-1/2 point-like particles

According to Einstein's mass-energy relation $E^2 = c^2(m_0^2 c^2 + p^2)$, the corresponding operator formalism for a free massless particle in the 2D spacetime is given by

$$(E^2 - c^2 P^2)|\Psi\rangle = 0 \tag{1}$$

There are two types of solutions to the above equation, the type-I solution with a commutative relation $[E, P] = 0$, and

$$E^2|\Psi\rangle = -E^2|\Psi\rangle, \quad P^2|\Psi\rangle = -P^2|\Psi\rangle. \tag{2}$$

which corresponds to a boson. For the type-II solution one has a non-commutative relation with $\{E, P\} = 0$, and

$$(E - cP)|\Psi\rangle = 0, \quad \text{or} \quad (E + cP)|\Psi\rangle = 0 \tag{3}$$

which corresponds to fermion.

For the type-II solution in 4D spacetime, one has the following set of criteria, i.e., the anti-commutative relation $\{E, P_k\} = 0, k = 1, 2, 3, \{P_i, P_j\} = 0$, if $i \neq j$, and

$$\begin{aligned}(E \mp cP)|\Psi\rangle &= 0, \\ P &= P_1 + P_2 + P_3.\end{aligned} \tag{4A}$$

Because $\{E, P_k\} = 0$ one has $\{E, P\} = EP + PE = 0$, and one can show

$$(E \mp c\ P)^2|\Psi\rangle = (E^2 + c^2 P^2 \mp c(EP + PE))|\Psi\rangle = (-E^2 + c^2 P^2)|\Psi\rangle = 0 \tag{4B}$$

We could assign the following 4x4 matrices to represent $E$, $P_k$ as

$$E = \boldsymbol{u}_0 E, \quad \boldsymbol{P}_k = \boldsymbol{u}_k P_k, \quad \{E, P_k\} = 0, \quad k = 1, 2, 3$$
$$\boldsymbol{u}_0 = \sigma_t \otimes I_2 = \begin{pmatrix} 0 & I_2 \\ -I_2 & 0 \end{pmatrix}, \boldsymbol{u}_k = \sigma_1 \otimes \sigma_k = \begin{pmatrix} 0 & \sigma_k \\ \sigma_k & 0 \end{pmatrix}. \tag{5A}$$

Owing to $\boldsymbol{u}_0^2 = -I$, $\boldsymbol{u}_k^2 = I$, one has $(E \mp cP) = 0$ and $\{E, P_k\} = 0$ which are the criteria for the type-II solution for the 4D case. The above operator assignment follows exactly the quaternion algebra[8]. Using the time and space-derivative representation for $E = \boldsymbol{u}_0 \hbar\, \partial/\partial t$, $\boldsymbol{P}_k = \boldsymbol{u}_k \hbar\, \partial/\partial x_k$ one can obtain from

$$\frac{\partial}{c\,\partial t} f(t,r) = \sigma \cdot \nabla f(t,r),$$
$$\frac{\partial}{c\,\partial t} g(t,r) = -\sigma \cdot \nabla g(t,r), \tag{5B}$$

and for $(E + cP)|\Psi\rangle = 0$ with $E = -u_t \hbar\, \partial/\partial t$, $P_k = u_k \hbar\, \partial/\partial x_k$ one has

$$\frac{\partial}{c\,\partial t} f(t,r) = -\sigma \cdot \nabla f(t,r),$$
$$\frac{\partial}{c\,\partial t} g(t,r) = \sigma \cdot \nabla g(t,r). \tag{5C}$$

Equations (5B) and (5C) describe the real-value wave equations of a massless fermion with an opposite propagation direction.

To satisfy the constraints of $u_0^2 = -I$, $u_k^2 = I$, one could also assign $E$, $P_k$ to different 4x4 matrices from those used previously. For example, one could assign

$$E = u_0 E,\ P_k = u_k P_k,\ \{E, P_k\} = 0,\ k = 1,2,3$$
$$u_0 = \sigma_t \otimes I_2 = \begin{pmatrix} 0 & I_2 \\ -I_2 & 0 \end{pmatrix},\ u_k = \sigma_3 \otimes \sigma_k = \begin{pmatrix} \sigma_k & 0 \\ 0 & -\sigma_k \end{pmatrix} \tag{6A}$$

$$\sigma_t = \begin{pmatrix} 0 & 1 \\ -1 & 0 \end{pmatrix},\ \sigma_1 = \begin{pmatrix} 0 & 1 \\ 1 & 0 \end{pmatrix},\ \sigma_2 = \begin{pmatrix} 0 & -i \\ i & 0 \end{pmatrix},\ \sigma_3 = \begin{pmatrix} 1 & 0 \\ 0 & -1 \end{pmatrix}.$$

These 4x4 tensor-product matrices can be related to Dirac's $\alpha_k$ and $\beta$ matrices or gamma matrices. Equations (5B) and (5C) become

$$\frac{\partial}{c\,\partial t} f(t,r) = -\sigma \cdot \nabla g(t,r),$$
$$\frac{\partial}{c\,\partial t} g(t,r) = \sigma \cdot \nabla f(t,r), \tag{6B}$$

and

$$\frac{\partial}{c\,\partial t} f(t,r) = \sigma \cdot \nabla g(t,r),$$
$$\frac{\partial}{c\,\partial t} g(t,r) = -\sigma \cdot \nabla f(t,r). \tag{6C}$$

All these coupled wave equations in Eqs.(5C), (6A), (6B) and (6C) involving 1st-order derivatives in time and space can be shown to become

$$-\frac{\partial^2}{c^2 \partial t^2} f(t,r) + \nabla^2 f(t,r) = 0,\quad -\frac{\partial^2}{c^2 \partial t^2} g(t,r) + \nabla^2 g(t,r) = 0. \tag{6D}$$

The above equations are 2nd-order partial differential equations for the wave equation of a massless particle which satisfies Einstein's relativistic mass-energy relation of $E^2 = c^2 p^2$.

## 2.2. Spin-1/2 point-like particles with a rest mass

Now, we extend our model description to a free relativistic massive particle, Eq. (4) becomes

$$\left(E^2 - c^2\left(P_1^2 + P_2^2 + P_3^2\right)\right)|\Psi\rangle = m_0^2 c^4 |\Psi\rangle, \qquad (7)$$

where the operator for the rest-mass energy $m_0 c^2$ is defined as $P_4 c$. For the type-II fermionic solution, one has the anti-commutative relations $\{P_i, P_j\} = 0$, if $i \neq j$, and

$$\begin{aligned}(E \mp c\mathbf{P})|\Psi\rangle &= \pm m_0 c^2 |\Psi\rangle, \\ \mathbf{P} &= \mathbf{P}_1 + \mathbf{P}_2 + \mathbf{P}_3.\end{aligned} \qquad (8\mathrm{A})$$

Because of the anti-commutation relations one can show

$$(E \mp c\mathbf{P})^2 |\Psi\rangle = \left(E^2 + c^2 \mathbf{P}^2 \mp c(E\mathbf{P} + \mathbf{P}E.)\right)|\Psi\rangle = \left(-E^2 + c^2 P^2\right)|\Psi\rangle = m_0^2 c^4 |\Psi\rangle \qquad (8\mathrm{B})$$

We could assign the following 4x4 matrices to represent $E, P_k$ as

$$\mathbf{E} = \mathbf{u}_0 E, \ \mathbf{P}_k = \mathbf{u}_k P_k, \ \{\mathbf{E}, \mathbf{P}_k\} = 0, k = 1, 2, 3,$$

$$\mathbf{u}_0 = \sigma_3 \otimes \mathbf{I}_2 = \begin{pmatrix} \mathbf{I}_2 & 0 \\ 0 & -\mathbf{I}_2 \end{pmatrix} \qquad (8\mathrm{C})$$

$$\mathbf{u}_k = \sigma_t \sigma_k = \begin{pmatrix} 0 & \sigma_k \\ -\sigma_k & 0 \end{pmatrix}, \ k = 1, 2, 3.$$

Using the time and space-derivative representation for $E = \mathbf{u}_0 \partial/\partial t, P_k = \mathbf{u}_k \partial/\partial x_k$ by extending Eqs. (8A) and (8B) for a massless fermion to the present case, one obtains the following linearized 1st-order rate equations

$$\begin{aligned}\frac{\partial}{c\,\partial t} f(t,\mathbf{r}) &= \boldsymbol{\sigma} \cdot \nabla g(t,\mathbf{r}) \pm (m_o c/\hbar) g(t,\mathbf{r}), \\ \frac{\partial}{c\,\partial t} g(t,\mathbf{r}) &= -\boldsymbol{\sigma} \cdot \nabla f(t,\mathbf{r}) + (m_o c/\hbar) f(t,\mathbf{r}),\end{aligned} \qquad (9\mathrm{A})$$

and

$$\begin{aligned}\frac{\partial}{c\,\partial t} f(t,\mathbf{r}) &= -\boldsymbol{\sigma} \cdot \nabla g(t,\mathbf{r}) \pm (m_o c/\hbar) g(t,\mathbf{r}), \\ \frac{\partial}{c\,\partial t} g(t,\mathbf{r}) &= +\boldsymbol{\sigma} \cdot \nabla f(t,\mathbf{r}) \pm (m_o c/\hbar) f(t,\mathbf{r}).\end{aligned} \qquad (9\mathrm{B})$$

The above waved equations are equivalent to Dirac's equation for a relativistic electron [14]. In covariant-form Dirac equation, one has $\left(i\hbar\gamma^\mu \partial_\mu - m_0 c\right)\Psi = 0$, where $p_k = -i\hbar\partial_k$, $p_0 = i\hbar\partial_0$, $\gamma^k = \sigma_1 \otimes \sigma_k$, and $\gamma^0 = \sigma_3 \otimes \mathbf{I}_2$. One can show that the coupled

wave equations in Equations (9A) and (9B) involving the 1$^{st}$-order derivatives in time and space lead to

$$-\frac{\partial^2}{c^2 \partial t^2} f(t,\mathbf{r}) + \nabla^2 f(t,\mathbf{r}) = (m_o c/\hbar)^2 f(t,\mathbf{r}),$$
$$-\frac{\partial^2}{c^2 \partial t^2} g(t,\mathbf{r}) + \nabla^2 g(t,\mathbf{r}) = (m_o c/\hbar)^2 g(t,\mathbf{r}).$$
(9C)

The above equations become the well-known Klein-Gordon for a massive scalar particle wave equation which satisfies Einstein's relativistic mass-energy relation of $E^2 = m_0^2 c^4 + c^2 p^2$.

### 2.3. Octonion electron with an internal dynamics structure

To generalize Dirac's equation for a spin-1/2 point-size fermion to a lepton with an internal structure, we need to include internal degrees of freedom by adding three more spatial momentum components $\boldsymbol{Q}_k, k = 1,2,3,$ to describe the internal dynamics with respect to the reference frame of a particle's center of mass. To satisfy $\{E, \boldsymbol{P}_k\} = \{E, \boldsymbol{Q}_k\} = 0, \{\boldsymbol{P}_i, \boldsymbol{Q}_j\} = 0,$ we have

$$(\mathbf{E} - c\mathbf{P})|\Psi\rangle = \pm m_0 c^2 |\Psi\rangle, \text{ or } (\mathbf{E} + c\mathbf{P})|\Psi\rangle = \pm m_0 c^2 |\Psi\rangle,$$
$$\mathbf{P} = \sum_{k=1}^{3} (\mathbf{P}_k + \mathbf{Q}_k).$$
(10)

For the description of massive spin-1/2 particle, unlike the previous section for bosons, we need to have a different assignment for the energy operator for $E$ so that $\{E, \boldsymbol{P}_k\} = \{E, \boldsymbol{Q}_k\} = 0,$ and $\{E, \boldsymbol{P}_4\} = 0.$ To satisfy these anti-commutative relationship, we consider octonion operators. Before we propose the assignments of these operators to describe leptons, we need to discuss some preliminary remarks about octonion algebra [17, 18]. At first, any element $x$ and $\bar{x}$, as its conjugate, in the octonion algebra can be expressed in terms of the identity operator $\boldsymbol{e}_0$ and seven other hyper-complex octonion unit operators $\boldsymbol{e}_k$ as

$$x = x_0 \mathbf{e}_0 + X, \quad \bar{x} = x_0 \mathbf{e}_0 - X, \quad X = \sum_{k=1}^{7} x_k \mathbf{e}_k,$$
(9A)

where $\boldsymbol{e}_k$ is the unit octonion operator which satisfies the anti-commutative relation

$\{e_i, e_j\} = 0$, $i \neq j$ for any pair of indices except $e_0$. The multiplication rules for these octonion operators are given in Fig. 1.

| $e_i \, e_j$ | $e_0$ | $e_1$ | $e_2$ | $e_3$ | $e_4$ | $e_5$ | $e_6$ | $e_7$ |
|---|---|---|---|---|---|---|---|---|
| $e_0$ | $e_0$ | $e_1$ | $e_2$ | $e_3$ | $e_4$ | $e_5$ | $e_6$ | $e_7$ |
| $e_1$ | $e_1$ | $-e_0$ | $e_3$ | $-e_2$ | $e_5$ | $-e_4$ | $-e_7$ | $e_6$ |
| $e_2$ | $e_2$ | $-e_3$ | $-e_0$ | $e_1$ | $e_6$ | $e_7$ | $-e_4$ | $-e_5$ |
| $e_3$ | $e_3$ | $e_2$ | $-e_1$ | $-e_0$ | $e_7$ | $-e_6$ | $e_5$ | $-e_4$ |
| $e_4$ | $e_4$ | $-e_5$ | $-e_6$ | $-e_7$ | $-e_0$ | $e_1$ | $e_2$ | $e_3$ |
| $e_5$ | $e_5$ | $e_4$ | $-e_7$ | $e_6$ | $-e_1$ | $-e_0$ | $-e_3$ | $e_2$ |
| $e_6$ | $e_6$ | $e_7$ | $e_4$ | $-e_5$ | $-e_2$ | $e_3$ | $-e_0$ | $-e_1$ |
| $e_7$ | $e_7$ | $-e_6$ | $e_5$ | $e_4$ | $-e_3$ | $-e_2$ | $e_1$ | $-e_0$ |

Fig. 1. The multiplication table for the eight unit octonions. According to the above multiplication rules, other than the identity operators, all other seven operators anti-commute with each other, and the multiplications are not always associative. The 8x8 arrays in the table are color-coded to illustrate several triads of cyclic operators.

According to the multiplication rules for any two elements $x$ and $y$ in the octonion algebra, their product can be expressed as

$$x \circ y = x_0 y_0 + x_0 \mathbf{e}_0 + x_0 \mathbf{Y} + y_0 \mathbf{X} - \mathbf{X} \cdot \mathbf{Y} + \mathbf{X} \times \mathbf{Y}, \tag{9B}$$

where $\mathbf{X} \cdot \mathbf{Y}$ is the inner product and $\mathbf{X} \times \mathbf{Y}$ is the cross product between $\mathbf{X}$ and $\mathbf{Y}$. The norm of an octonion $x$ is $x \circ \bar{x} = \bar{x} \circ x = \sum_{n=0}^{7} x_n^2$.

To construct the matrix representation for an octonion operator, one first defines

$$\mathbf{u}_0 = (\mathbf{e}_0 + i\mathbf{e}_4)/2 = \begin{pmatrix} 0 & 0 \\ 0 & 1 \end{pmatrix}, \quad \mathbf{u}_0^* = (\mathbf{e}_0 - i\mathbf{e}_4)/2 = \begin{pmatrix} 1 & 0 \\ 0 & 0 \end{pmatrix},$$

$$\mathbf{u}_k = (\mathbf{e}_k + i\mathbf{e}_{k+4})/2 = \begin{pmatrix} 0 & 0 \\ \mathbf{e}_k & 0 \end{pmatrix}, \quad \mathbf{u}_k^* = (\mathbf{e}_k - i\mathbf{e}_{k+4})/2 = \begin{pmatrix} 0 & -\mathbf{e}_k \\ 0 & 0 \end{pmatrix} \tag{9C}$$

$$\mathbf{e}_0 = \begin{pmatrix} 1 & 0 \\ 0 & 1 \end{pmatrix}, \mathbf{e}_1 = -i\begin{pmatrix} 0 & 1 \\ 1 & 0 \end{pmatrix}, \mathbf{e}_2 = -i\begin{pmatrix} 0 & -i \\ i & 0 \end{pmatrix}, \mathbf{e}_3 = -i\begin{pmatrix} 1 & 0 \\ 0 & -1 \end{pmatrix}.$$

According to Zorn's scheme for matrix construction to satisfy the octonion multiplication rules in the octonion algebra. Let us consider $\mathbf{Z} = \sum_{k=0}^{7} z_0 \mathbf{e}_k$, and and define

$$a = z_0 + iz_4, \ b = z_0 - iz_4, \ C_k = -z_k + iz_{k+4}, D_k = z_k + i z_{k+4},$$
$$a' = z_0' + iz_4', \ b' = z_0' - iz_4', \ C_k' = -z_k' + iz_{k+4}', D_k' = z_k' + i z_{k+4}', \tag{10A}$$

According to Zorn's vector-matrix representation [17, 18], one has the following multiplication rules to guarantee the constraints in Eq. (12B),

$$\mathbf{Z} = \begin{pmatrix} a & \mathbf{C} \\ \mathbf{D} & b \end{pmatrix}, \mathbf{Z}' = \begin{pmatrix} a' & \mathbf{C}' \\ \mathbf{D}' & b' \end{pmatrix}$$

$$\begin{pmatrix} a & \mathbf{C} \\ \mathbf{D} & b \end{pmatrix} \circ \begin{pmatrix} a' & \mathbf{C}' \\ \mathbf{D}' & b' \end{pmatrix} = \begin{pmatrix} aa' + \mathbf{C} \cdot \mathbf{D}' & a\mathbf{C}' + b'\mathbf{C} - \mathbf{D} \times \mathbf{D}' \\ a'\mathbf{D} + b\mathbf{D}' + \mathbf{C} \times \mathbf{C}' & bb' + \mathbf{D} \cdot \mathbf{C}' \end{pmatrix}. \tag{10B}$$

And, according to Eq. (13B) and (13A) one obtains

$$\overline{\mathbf{Z}} = \begin{pmatrix} b & -\mathbf{C} \\ -\mathbf{D} & a \end{pmatrix},$$

$$\mathbf{Z} \circ \overline{\mathbf{Z}} = \begin{pmatrix} ab - \mathbf{C} \cdot \mathbf{D} & a\mathbf{C} - a\mathbf{C} + \mathbf{D} \times \mathbf{D} \\ b\mathbf{D} - b\mathbf{D} - \mathbf{C} \times \mathbf{C} & ba - \mathbf{D} \cdot \mathbf{C} \end{pmatrix} = \begin{pmatrix} ab - \mathbf{C} \cdot \mathbf{D} & 0 \\ 0 & ba - \mathbf{D} \cdot \mathbf{C} \end{pmatrix} \tag{10C}$$

$$== (ab - (\mathbf{C} \cdot \mathbf{D} + \mathbf{D} \cdot \mathbf{C})/2)\mathbf{I} - \sigma_3 \otimes (\mathbf{C} \cdot \mathbf{D} - \mathbf{D} \cdot \mathbf{C})/2$$

For octonion algebra, $\mathbf{C} \cdot \mathbf{D} = \mathbf{D} \cdot \mathbf{C}$ and

$$\mathbf{Z} \circ \overline{\mathbf{Z}} == (ab - \mathbf{C} \cdot \mathbf{D})\mathbf{I} = \sum_{k=0}^{7} z_k^2 \mathbf{I} \tag{10D}$$

Using the natural unit of $c = \hbar = 1$, we define an element $\mathbf{Z}$ in the octonion algebra as

$$\mathbf{E} = E\mathbf{e}_4 \ \mathbf{M} = m_0 \mathbf{e}_0$$
$$\alpha = z_0 + i z_4, \quad \beta = z_0 - i z_4 \tag{11A}$$
$$\mathbf{Z} = \sum_{k=0}^{7} z_0 \mathbf{e}_k \mapsto \begin{pmatrix} a & -\mathbf{P} + i\mathbf{Q} \\ \mathbf{P} + i\mathbf{Q} & b \end{pmatrix} = \begin{pmatrix} z_0 + iz_4 & -z_k + iz_{k+4} \\ z_k + i z_{k+4} & z_0 - iz_4 \end{pmatrix},$$

where the symbol $\mapsto$ represents an isomorphism.

To represent leptons, we need to assign the appropriate operators to the energy and six generalized momentum components. Here, we consider $E = iE\sigma_3 \otimes I_2, M = m_0 I$, $\alpha = M + E, \beta = M - E$, and

$$Z \mapsto \begin{pmatrix} a & -\mathbf{P}+i\mathbf{Q} \\ \mathbf{P}+i\mathbf{Q} & b \end{pmatrix} \qquad (11B)$$
$$= m_0 I + iE\sigma_3 \otimes I_2 - \sigma_t \otimes \mathbf{P} + i\sigma_1 \otimes \mathbf{Q}$$

and

$$\mathbf{P} = \sum_{k=1}^{3} P_k \sigma_k \mapsto \sum_{k=1}^{3} z_k \mathbf{e}_k$$
$$\mathbf{Q} = \sum_{k=1}^{3} Q_k \sigma_k \mapsto \sum_{k=1}^{3} z_{k+4} \mathbf{e}_{k+4} \qquad (11C)$$
$$\sigma_i \sigma_j = i\varepsilon_{ijk}\sigma_k + \delta_{ij} I_2, \quad \mathbf{e}_i \mathbf{e}_i = -\mathbf{e}_0, \quad \mathbf{e}_i, \mathbf{e}_{j+4} = -\mathbf{e}_{j+4} \mathbf{e}_i.$$

Following Zorn's multiplication rules, the norm of the octonion is given by

$$\mathbf{Z} \circ \overline{\mathbf{Z}} = \alpha\beta - (-\mathbf{P}+i\mathbf{Q}) \cdot (\mathbf{P}+i\mathbf{Q})$$
$$= m_0^2 - E^2 - (-z_k + iz_{k+4})(z_k + iz_{k+4}) \qquad (11D)$$
$$= m_0^2 - E^2 + \sum_k P_k^2 + \sum_k Q_k^2.$$

With $\mathbf{Z} \circ \overline{\mathbf{Z}} = 0$, one obtains Einstein's relativistic mass-energy relation $E = \sqrt{m_0^2 + \sum_k A_k^2 + \sum_k B_k^2} \equiv \sqrt{m_{0,\text{eff}}^2 + \sum_k P_k^2}$, where $\sum_k P_k^2$ is the kinetic energy in the laboratory frame, $\sum_k Q_k^2$ the internal kinetic energy with respect to the frame of the particle's center of mass, and $m_{0,\text{eff}} = \sqrt{m_0^2 + \sum_k Q_k^2}$ the effective rest-mass by including the kinetic energy from the particle's internal structural dynamics.

The reasons for making the above operator assignment are explained in the following. We chose $(\sigma_1 + \sigma_t)/2$ in $-\mathbf{P} + i\mathbf{Q}$, $(\sigma_1 - \sigma_t)/2$ in $\mathbf{P} + i\mathbf{Q}$, and $\sigma_3$ in $E$ is to ensure them to be anti-commutes with each other. Three operators $\mathbf{P}_k = P_k \sigma_k, k = 1,2,3$, describes the motion of the particle's center of mass with respect to the laboratory frame, and $\mathbf{Q}_k = Q_k \sigma_k$ describes the particle's internal structural dynamics with respect to the rest frame of its center of mass. If one se $B_k = 0$, i.e., assuming the particle with no internal degrees of freedom, Equations (11A-11D) become the usual Dirac equation for a point-like electron. In short, the use of the octonion operators automatically invokes an internal degree of freedom.

**2.4. Sedenion algebra and octonion sub-algebra for three lepton generations**

In the previous section, the 2x2 Pauli matrices $\sigma_k$ were used in constructing both $\mathbf{P}_k = P_k \sigma_k$ and $\mathbf{Q}_k = Q_k \sigma_k$, such an assignment only leads to one type of lepton. To accommodate three generations of leptons as proposed in the Standard Model, i.e., an electron, muon and tau, we need to consider the sedenion algebra [20-23] with sixteen unit operators, which is an extension of the octonion algebra with eight operators. The multiplication rules for sixteen sedenion unit operators are given in Fig. 2.

Fig. 2. The multiplication table for the unit sedenions containing $\{e_k, k=0,1,2,...,15\}$, which are denoted by $\{\mathbf{I}, \mathbf{\Gamma}_1, \mathbf{\Gamma}_2, \mathbf{\Gamma}_3, \mathbf{\Theta}_1, \mathbf{U}_1, \mathbf{U}_2, \mathbf{U}_3, \mathbf{\Theta}_2, \mathbf{V}_1, \mathbf{V}_2, \mathbf{V}_3, \mathbf{\Theta}_3, \mathbf{W}_1, \mathbf{W}_2, \mathbf{W}_3\}$ sequentially. The table is color-coded to better show five different triads such as $\mathbf{U}_k, \mathbf{V}_k, \mathbf{W}_k, \mathbf{\Gamma}_k, \mathbf{\Theta}_k, k=1,2,3$, which follow the same cyclic multiplication rule similar as Pauli's SU(2) spinors.

According to the multiplication rules shown in Fig. 2, there are three types of octonion algebras as the sub-algebras of the sedenion, namely, $\{\mathbf{I}, \mathbf{\Gamma}_k, \mathbf{\Theta}_1 \cdot \mathbf{U}_k\}$, $\{\mathbf{I}, \mathbf{\Gamma}_k, \mathbf{\Theta}_2 \cdot \mathbf{V}_k\}$, and $\{\mathbf{I}, \mathbf{\Gamma}_k, \mathbf{\Theta}_3 \cdot \mathbf{W}_k\}$, where k = 1, 2, 3. These three octonion algebras represent three generation of massive leptons, i.e., electron, muon and tau. The multiplication rules and the corresponding tables for three types of octonion algebra are illustrated in the following sequentially.

| | I | $\Gamma_1$ | $\Gamma_2$ | $\Gamma_3$ | $\Theta_1$ | $U_1$ | $U_2$ | $U_3$ |
|---|---|---|---|---|---|---|---|---|
| $\Gamma_1$ | -I | $\Gamma_3$ | $-\Gamma_2$ | $U_1$ | $-\Theta_1$ | $-U_3$ | $U_2$ |
| $\Gamma_2$ | $-\Gamma_3$ | -I | $\Gamma_1$ | $U_2$ | $U_3$ | $-\Theta_1$ | $-U_1$ |
| $\Gamma_3$ | $\Gamma_2$ | $-\Gamma_1$ | -I | $U_3$ | $-U_2$ | $U_1$ | $-\Theta_1$ |
| $\Theta_1$ | $-U_1$ | $-U_2$ | $-U_3$ | -I | $\Gamma_1$ | $\Gamma_2$ | $\Gamma_3$ |
| $U_1$ | $\Theta_1$ | $-U_3$ | $U_2$ | $-\Gamma_1$ | -I | $-\Gamma_3$ | $\Gamma_2$ |
| $U_2$ | $U_3$ | $\Theta_1$ | $-U_1$ | $-\Gamma_2$ | $\Gamma_3$ | -I | $-\Gamma_1$ |
| $U_3$ | $-U_2$ | $U_1$ | $\Theta_1$ | $-\Gamma_3$ | $-\Gamma_2$ | $\Gamma_1$ | -I |

Fig. 3. The color-coded multiplication rules for the octonions of $\{I, \Gamma_1, \Gamma_2, \Gamma_3, \Theta_1, U_1, U_2, U_3\}$

| | $e_0$ | $e_1$ | $e_2$ | $e_3$ | $e_8$ | $e_9$ | $e_{10}$ | $e_{11}$ |
|---|---|---|---|---|---|---|---|---|
| $e_0$ | I | $\Gamma_1$ | $\Gamma_2$ | $\Gamma_3$ | $\Theta_2$ | $V_1$ | $V_2$ | $V_3$ |
| $e_1$ | $\Gamma_1$ | -I | $\Gamma_3$ | $-\Gamma_2$ | $V_1$ | $-\Theta_2$ | $-V_3$ | $V_2$ |
| $e_2$ | $\Gamma_2$ | $-\Gamma_3$ | -I | $\Gamma_1$ | $V_2$ | $U_3$ | $-\Theta_2$ | $-V_1$ |
| $e_3$ | $\Gamma_3$ | $\Gamma_2$ | $-\Gamma_1$ | -I | $V_3$ | $-V_2$ | $V_1$ | $-\Theta_2$ |
| $e_8$ | $\Theta_2$ | $-V_1$ | $-V_2$ | $-V_3$ | -I | $\Gamma_1$ | $\Gamma_2$ | 3 |
| $e_9$ | $V_1$ | $\Theta_2$ | $-V_3$ | $V_2$ | $-\Gamma_1$ | -I | $-\Gamma_3$ | $\Gamma_2$ |
| $e_{10}$ | $V_2$ | $U_3$ | $\Theta_2$ | $-V_1$ | $-\Gamma_2$ | $\Gamma_3$ | -I | $-\Gamma_1$ |
| $e_{11}$ | $V_3$ | $-V_2$ | $V_1$ | $\Theta_2$ | $-\Gamma_3$ | $-\Gamma_2$ | $\Gamma_1$ | -I |

Fig. 4. The color-coded multiplication rules for the octonions of $\{I, \Gamma_1, \Gamma_2, \Gamma_3, \Theta_2, V_1, V_2, V_3\}$

| | **I** | **Γ₁** | **Γ₂** | **Γ₃** | **Θ₃** | **W₁** | **W₂** | **W₃** |
|---|---|---|---|---|---|---|---|---|
| **Γ₁** | -I | Γ₃ | -Γ₂ | -W₁ | Θ₃ | W₃ | -W₂ | |
| **Γ₂** | -Γ₃ | -I | Γ₁ | -W₂ | -W₃ | Θ₃ | W₁ | |
| **Γ₃** | Γ₂ | -Γ₁ | -I | -W₃ | W₂ | -W₁ | Θ₃ | |
| **Θ₃** | W₁ | W₂ | W₃ | -I | −Γ₁ | −Γ₂ | −Γ₃ | |
| **W₁** | -Θ₃ | W₃ | -W₂ | Γ₁ | -I | Γ₃ | −Γ₂ | |
| **W₂** | -W₃ | -Θ₃ | W₁ | Γ₂ | −Γ₃ | -I | Γ₁ | |
| **W₃** | W₂ | -W₁ | -Θ₃ | Γ₃ | Γ₂ | −Γ₁ | -I | |

Fig. 5. The color-coded multiplication rules for the octonions of $\{\mathbf{I}, \mathbf{\Gamma}_1, \mathbf{\Gamma}_2, \mathbf{\Gamma}_3, \mathbf{\Theta}_3, \mathbf{W}_1, \mathbf{W}_2, \mathbf{W}_3\}$

To represent an electron using the first set of octonions $\{\mathbf{I}, \mathbf{\Gamma}_1, \mathbf{\Gamma}_2, \mathbf{\Gamma}_3, \mathbf{\Theta}_1, \mathbf{U}_1, \mathbf{U}_2, \mathbf{U}_3\}$, similar to the earlier discussion in the last section, we consider $a = z_0 + i z_4$, $b = z_0 - i z_4$, $C_k = -z_k + i z_{k+4}$, $D_k = z_k + i z_{k+4}$, and

$$\mathbf{E} = E\mathbf{\Theta}_1, \quad \mathbf{M} = m_0\mathbf{I}, \quad \mathbf{P} = \sum_{k=1}^{3} P_k \mathbf{\Gamma}_k, \quad \mathbf{Q} = \sum_{k=1}^{3} Q_{uk} \mathbf{U}_k$$

$$\alpha = z_0 + i z_4, \quad \beta = z_0 - i z_4 \tag{12A}$$

$$\mathbf{Z} = \sum_{k=0}^{7} z_0 \mathbf{e}_k \mapsto \begin{pmatrix} a & -\mathbf{P} + i\mathbf{Q}_u \\ \mathbf{P} + i\mathbf{Q}_u & b \end{pmatrix} = \begin{pmatrix} z_0 + iz_4 & -z_k + iz_{k+4} \\ z_k + i z_{k+4} & z_0 - iz_4 \end{pmatrix},$$

We obtain

$$\mathbf{Z} \circ \mathbf{Z} = m_0^2 - E^2 + \sum_k P_k^2 + \sum_k Q_{uk}^2 = 0, \tag{12B}$$

which leads to Einstein's mass-energy relation for an electron with an internal mass energy $m_{u,\text{eff}} = \sqrt{m_0^2 + \sum_k Q_{uk}^2}$ due to its internal structural dynamics. Similar to the U-type lepton, for the two other **V**-type octonion algebra $\{\mathbf{I}, \mathbf{\Gamma}_1, \mathbf{\Gamma}_2, \mathbf{\Gamma}_3, \mathbf{\Theta}_2, \mathbf{V}_1, \mathbf{V}_2, \mathbf{V}_3\}$ one has

$$\mathbf{E} = E\mathbf{\Theta}_2, \quad \mathbf{M} = m_0\mathbf{I}, \quad \mathbf{P} = \sum_k P_k \mathbf{\Gamma}_k, \quad \mathbf{Q} = \sum_k Q_{vk} \mathbf{V}_k, \tag{12C}$$

For the **W**-type octonion algebra $\{\mathbf{I}, \mathbf{\Gamma}_1, \mathbf{\Gamma}_2, \mathbf{\Gamma}_3, \mathbf{\Theta}_3, \mathbf{W}_1, \mathbf{W}_2, \mathbf{W}_3\}$ one has

$$\mathbf{E} = E\mathbf{\Theta}_3, \quad \mathbf{M} = m_0\mathbf{I}, \quad \mathbf{P} = \sum_k P_k \mathbf{\Gamma}_k \quad \mathbf{Q} = \sum_k Q_{wk} \mathbf{W}_k. \tag{12D}$$

From Eqs. (12C) and (12D) one obtains the effective rest masses

$m_{v,\text{eff}} = \sqrt{m_0^2 + \sum_k Q_{vk}^2}$ and $m_{w,\text{eff}} = \sqrt{m_0^2 + \sum_k Q_{wk}^2}$, respectively. Therefore, the three sub-algebras of the sedenion algebra correspond to three generations of the charged leptons, i.e., the electron, muon and tau.

For each type of lepton, we can show in the following that the octonion algebra is isomorphic to Clifford algebra $C\ell(6)$. We first define a set of creation and annihilation operators $\alpha_k^+$ and $\alpha_k$ which satisfy the anti-commutation relations of a fermion as

$$\alpha_1 = (-e_6 + ie_5)/2, \quad \alpha_2 = (-e_3 + ie_1)/2, \quad \alpha_3 = (-e_7 + ie_2)/2$$
$$\{\alpha_i, \alpha_j\} = \{\alpha_i^+, \alpha_j^+\} = 0, \quad \{\alpha_i, \alpha_j^+\} = \delta_{ij} \tag{13A}$$

Or, using the notation of $\{\mathbf{I}, \mathbf{\Gamma}_k, \mathbf{\Theta}_1, \mathbf{U}_k\}$, for the first generation

$$\alpha_1 = (-\mathbf{U}_2 + i\mathbf{U}_1)/2, \alpha_2 = (-\mathbf{\Gamma}_3 + i\mathbf{\Gamma}_1)/2, \alpha_3 = (-\mathbf{U}_3 + i\mathbf{\Gamma}_2)/2$$
$$\mathbf{U}_k^+ = -\mathbf{U}_k, \mathbf{V}_k^+ = -\mathbf{V}_k, \mathbf{W}_k^+ = -\mathbf{W}_k, (e_k^+ = -e_k, k=1,2,3)$$
$$a_1^+ = (\mathbf{U}_2 + i\mathbf{U}_1)/2, a_2^+ = (\mathbf{\Gamma}_3 + i\mathbf{\Gamma}_1)/2, a_3^+ = (\mathbf{U}_3 + i\mathbf{\Gamma}_2)/2 \tag{13B}$$
$$\{\alpha_k, \alpha_k\} = \{a_k^+, a_k^+\} = 0,$$
$$\{\alpha_i, \alpha_j^+\} = \delta_{ij}\mathbf{I}.$$

One can pair up these creation and annihilation operators, $|i\rangle\langle j| \equiv \alpha_i^+ \alpha_j$, to construct the lambda matrices $\Lambda_k$ of SU(3) generators as

$$\Lambda_1 = |2\rangle\langle 1| + |1\rangle\langle 2| = i(\mathbf{U}_3 - \mathbf{U}_2)/2$$
$$\Lambda_2 = -i|1\rangle\langle 2| + i|2\rangle\langle 1| = -i(\mathbf{U}_1 - \mathbf{\Theta}_1)/2$$
$$\Lambda_3 = |1\rangle\langle 1| - |2\rangle\langle 2| = i(\mathbf{\Gamma}_3 - \mathbf{\Gamma}_2)/2$$
$$\Lambda_4 = |1\rangle\langle 3| + |3\rangle\langle 1| = i(\mathbf{\Theta}_1 - \mathbf{\Gamma}_2)/2$$
$$\Lambda_5 = -i|1\rangle\langle 3| + i|3\rangle\langle 1| = -i(\mathbf{\Gamma}_1 + \mathbf{U}_3)/2 \tag{13C}$$
$$\Lambda_6 = |2\rangle\langle 3| + |3\rangle\langle 2| = -i(\mathbf{\Gamma}_1 + \mathbf{U}_2)/4$$
$$\Lambda_7 = -i|2\rangle\langle 3| + i|3\rangle\langle 2| = -i(\mathbf{\Theta}_1 - \mathbf{\Gamma}_3)/2$$
$$\Lambda_8 = (|1\rangle\langle 1| + |2\rangle\langle 2| - 2|3\rangle\langle 3|)/\sqrt{3} = i(\mathbf{\Gamma}_3 + \mathbf{\Gamma}_2 - 2\mathbf{U}_2)/2.$$

One can see from the above the isomorphism between the Clifford algebra $C\ell(6)$ and the SU(3). Together with the U-type, V-type and W-type Clifford algebras as the three sub-algebra of the sedenion algebra lead to three generations of quarks.

## 2.5. Three generations of neutrinos, their masses and the flavor-mixing mechanism

We have shown in the last section three different octonion algebras as the sub-algebras of the sixteen-dimensional sedenion algebra for the electron, muon and tau. Here, we present different operator assignments for the three generations of neutrinos in the Standard Model and explain why they have almost vanishingly small rest masses, and the causes for their flavor-mixing and mass oscillations emerge.

Fig. 6. The multiplication table for the unit sedenions, but arranged according to the sequence of $\{I, U_1, U_2, U_3, V_1, V_2, V_3, W_1, W_2, W_3, \Gamma_1, \Gamma_2, \Gamma_3, \Theta_1, \Theta_2, \Theta_3\}$. The table is color-coded into five different triad domains, representing the sub-algebra structures.

Here we provide the explanations to the origins of the smallness of the mass for three generations of the neutrinos and their mass oscillations induced by flavor-mixing. We can make the following assignment $C = -P + iQ$, $D = P + iQ$. According to Eq. (10C) we obtain $Z \circ \overline{Z} = (ab - (P \cdot P + Q \cdot Q))I + i\sigma_3 \otimes [P, Q]$, and for three generations of the neutrinos, we have

1) $\mathbf{E} = iE\Theta_1$, $\mathbf{M} = m_0\mathbf{I}$, $\mathbf{P} = \sum_{k=1}^{3} P_k \Gamma_k$, $\mathbf{Q} = \sum_{k=1}^{3} Q_k (\mathbf{V}_k - i\mathbf{W}_k)/\sqrt{2}$,

$\mathbf{Z} \circ \overline{\mathbf{Z}}$

$$= m_o^2 - E^2 + \sum_{k=1}^{3} P_k^2 - i\sqrt{2}\sigma_3 \otimes \left( (\Theta_2 + \Theta_3) \sum_{k=1}^{3} P_k Q_k + \sum_{i,j=1}^{3} \varepsilon_{ijk} P_i Q_j (\mathbf{V}_k + i\mathbf{W}_k) \right)$$

$$= m_o^2 - E^2 + (\mathbf{P} \cdot \mathbf{P}) - i\sqrt{2}\sigma_3 \otimes \left( (\mathbf{P} \cdot \mathbf{Q})(\Theta_2 + i\Theta_3) + (\mathbf{P} \times \mathbf{Q}) \cdot (\mathbf{V} + i\mathbf{W}) \right)$$

(14A)

2) $\mathbf{E} = iE\Theta_2$, $\mathbf{M} = m_0\mathbf{I}$,, $\mathbf{P} = \sum_{k=1}^{3} P_k \Gamma_k$, $\mathbf{Q} = \sum_{k=1}^{3} Q_k (\mathbf{W}_k - i\mathbf{U}_k)$

$\mathbf{Z} \circ \overline{\mathbf{Z}} = \backslash$

$$= m_o^2 - E^2 + \sum_{k=1}^{3} P_k^2 + i\sqrt{2}\sigma_3 \otimes \left( (i\Theta_3 + \Theta_1) \sum_{k=1}^{3} P_k Q_k + \sum_{i,j=1}^{3} \varepsilon_{ijk} P_i Q_j (\mathbf{W}_k + i\mathbf{U}_k) \right)$$

$$= m_o^2 - E^2 + (\mathbf{P} \cdot \mathbf{P}) + i\sqrt{2}\sigma_3 \otimes \left( (\mathbf{P} \cdot \mathbf{Q})(\Theta_3 + i\Theta_1) + (\mathbf{P} \times \mathbf{Q}) \cdot (\mathbf{W} + i\mathbf{U}) \right)$$

(13B)

and

3) $\mathbf{E} = iE\Theta_3$, $\mathbf{M} = m_0\mathbf{I}$,, $\mathbf{P} = \sum_{k=1}^{3} P_k \Gamma_k$, $\mathbf{Q} = \sum_{k=1}^{3} Q_k (\mathbf{U}_k - i\mathbf{V}_k)$

$\mathbf{Z} \circ \overline{\mathbf{Z}} =$

$$= m_o^2 - E^2 + \sum_{k=1}^{3} P_k^2 - i\sqrt{2}\sigma_3 \otimes \left( (i\Theta_1 + \Theta_2) \sum_{k=1}^{3} P_k Q_k + \sum_{i,j=1}^{3} \varepsilon_{ijk} P_i Q_j (i\mathbf{U}_k + \mathbf{V}_k) \right)$$

$$= m_o^2 - E^2 + (\mathbf{P} \cdot \mathbf{P}) + i\sqrt{2}\sigma_3 \otimes \left( (\mathbf{P} \cdot \mathbf{Q})(\Theta_1 + i\Theta_2) + (\mathbf{P} \times \mathbf{Q}) \cdot (\mathbf{U} + i\mathbf{V}) \right)$$

(14C)

where we have used $[\Gamma_i, \mathbf{U}_j] = -2\delta_{ij}\Theta_1 - 2\varepsilon_{ijk}\mathbf{U}_k$, $[\Gamma_i, \mathbf{V}_j] = -2\delta_{ij}\Theta_2 - 2\varepsilon_{ijk}\mathbf{V}_k$, wand $[\Gamma_i, \mathbf{W}_j] = 2\delta_{ij}\Theta_3 + 2\varepsilon_{ijk}\mathbf{W}_k$..

The above assugbnebts represent three corresponding neutrinos of the electron, muon and tau. For the electron, muon and tau, the norm leads to the usual Einstein's mass-energy relation with $E = \sqrt{m_0^2 + \sum_k P_k^2 + \sum_k Q_k^2}$. However, for the neutrinos, their norm $\mathbf{Z} \circ \overline{\mathbf{Z}} = 0$ does not lead to Einstein's relativistic mass-energy relations, due to the presence of the last non-vanishing commutator term. Even if the last flavor-mixing term is missing, one would obtain $E = \sqrt{m_0^2 + \sum_k P_k^2}$ for a neutrino, which differs from that of an electron with $E = \sqrt{m_0^2 + \sum_k P_k^2 + \sum_k Q_k^2}$. The absence of the term $\sum_k Q_k^2$ for neutrinos leads to their almost vanishing rest mass, and would have no rest mass if $m_0 = 0$ in $E = \sqrt{m_0{}^2 + \sum_k P_k^2}$. The presence of the last term in $\mathbf{Z} \circ \overline{\mathbf{Z}}$ indicates that the Hamiltonian for each type of neutrino does not commute with the mass operator, which implies that mass is not an eigenvalue, and mass oscillations would occur, and flavor mixing among the three types of neutrinos would naturally arise based on our proposed model.

For the electrons, muon and tau in Equations (13A), (13B) and (13C) with an effective mass-energy of $\sqrt{m_0^2 + \sum_k Q_k^2}$, where $\sum_k Q_k^2$ is the additional rest mass-energy from the internal structural dynamics. In contrast, for neutrinos, there is no such a term of $\sum_k B_k^2$, therefore, the rest mass of a neutrinos can be zero or vanishingly small due to the absence of $\sum_k B_k^2$. This is the physical explanation to why the neutrinos have such a vanishingly small mass, unlike their counterpart massive leptons. Moreover, because the presence of the last term in all three types of neutrinos contain

$$\sigma_3 \otimes \left( (\boldsymbol{P} \cdot \boldsymbol{Q})(\Theta_2 + i\Theta_3) + (\boldsymbol{P} \times \boldsymbol{Q}) \cdot (\mathbf{V} + i\mathbf{W}) \right), \quad \sigma_3 \otimes \left( (\boldsymbol{P} \cdot \boldsymbol{Q})(\Theta_3 + i\Theta_1) + (\boldsymbol{P} \times \boldsymbol{Q}) \cdot (\mathbf{W} + i\mathbf{U}) \right)$$

and $\sigma_3 \otimes \left( (\boldsymbol{P} \cdot \boldsymbol{Q})(\Theta_1 + i\Theta_2) + (\boldsymbol{P} \times \boldsymbol{Q}) \cdot (\mathbf{U} + i\mathbf{V}) \right)$, and these operator terms behave like Pauli matrices to produce cyclic rotations for the effective mass operator. Thus, even if $m_0$ has a zero or very small value, the last term in Equations (13A-13C) can be regarded as the source for the flavor-mixing mechanism to induce mass oscillations. Our model, therefore, explains the physical reasons for why neutrinos have three generations as well, they have a vanishingly small mass, and these flavors of neutrinos exhibit mass oscillation behavior which have been described based on a phenomenological Pontecorvo-Maki-Nakagawa-Sataka model [23-24].

The corresponding wave equation for the octonion Hamiltonian represents a generalization of the conventional Dirac equation. For a point-like electron, the Dirac equation involves 1st-order derivatives of space and time to describe the motion of the particle's center of mass. In contrast, because of the presence of a lepton's internal structural dynamics, our model involves three additional momentum components to describe a particle's internal structural dynamics with respect to the reference of the particle's center of mass. According to Zorn's vector-matrix representation, the

corresponding wave equation for our model for each type of leptons described in Eq. (13A-13C), one has

1)
$$Z\begin{pmatrix}\Psi\\ \Phi\end{pmatrix} \mapsto \begin{pmatrix} m_0 - \Theta_1 \partial_t & -\Gamma \cdot \nabla_{\mathbf{x}} + i\mathbf{U} \cdot \nabla_{\mathbf{q}} \\ \Gamma \cdot \nabla_{\mathbf{x}} + i\mathbf{U} \cdot \nabla_{\mathbf{q}} & m_0 + \Theta_1 \partial_t \end{pmatrix}\begin{pmatrix}\Psi\\ \Phi\end{pmatrix} = 0,$$

2)
$$Z\begin{pmatrix}\Psi\\ \Phi\end{pmatrix} \mapsto \begin{pmatrix} m_0 - \Theta_2 \partial_t & -\Gamma \cdot \nabla_{\mathbf{x}} + i\mathbf{V} \cdot \nabla_{\mathbf{q}} \\ \Gamma \cdot \nabla_{\mathbf{x}} + i\mathbf{V} \cdot \nabla_{\mathbf{q}} & m_0 + \Theta_2 \partial_t \end{pmatrix}\begin{pmatrix}\Psi\\ \Phi\end{pmatrix} = 0, \quad (15)$$

3)
$$Z\begin{pmatrix}\Psi\\ \Phi\end{pmatrix} \mapsto \begin{pmatrix} m_0 - \Theta_3 \partial_t & -\Gamma \cdot \nabla_{\mathbf{x}} + i\mathbf{W} \cdot \nabla_{\mathbf{q}} \\ \Gamma \cdot \nabla_{\mathbf{x}} + i\mathbf{W} \cdot \nabla_{\mathbf{q}} & m_0 + \Theta_3 \partial_t \end{pmatrix}\begin{pmatrix}\Psi\\ \Phi\end{pmatrix} = 0,$$

where $\nabla_x$ and $\nabla_q$ represent the gradient for the motion of the center of mass and the particle's internal dynamics, respectively. For neutrinos, according to Equations (14A-14C), one has

1)
$$\begin{pmatrix} m_0 - \Theta_1 \partial_t & -\Gamma \cdot \nabla_{\mathbf{x}} + (\mathbf{V} - i\mathbf{W}) \cdot \nabla_{\mathbf{q}} \\ \Gamma \cdot \nabla_{\mathbf{x}} + (\mathbf{V} - i\mathbf{W}) \cdot \nabla_{\mathbf{q}} & m_0 + \Theta_1 \partial_t \end{pmatrix}\begin{pmatrix}\Psi\\ \Phi\end{pmatrix} = 0$$

2)
$$\begin{pmatrix} m_0 - \Theta_2 \partial_t & -\Gamma \cdot \nabla_{\mathbf{x}} + (\mathbf{W} - i\mathbf{U}) \cdot \nabla_{\mathbf{q}} \\ \Gamma \cdot \nabla_{\mathbf{x}} + (\mathbf{W} - i\mathbf{U}) \cdot \nabla_{\mathbf{q}} & m_0 + \Theta_2 \partial_t \end{pmatrix}\begin{pmatrix}\Psi\\ \Phi\end{pmatrix} = 0 \quad (16)$$

3)
$$\begin{pmatrix} m_0 - \Theta_3 \partial_t & -\Gamma \cdot \nabla_{\mathbf{x}} + (\mathbf{U} - i\mathbf{V}) \cdot \nabla_{\mathbf{q}} \\ \Gamma \cdot \nabla_{\mathbf{x}} + (\mathbf{U} - i\mathbf{V}) \cdot \nabla_{\mathbf{q}} & m_0 + \Theta_3 \partial_t \end{pmatrix}\begin{pmatrix}\Psi\\ \Phi\end{pmatrix} = 0.$$

The relativistic wave equations shown above represent our generalized Dirac equation for three generations of neutrinos, exhibiting flavor-mixing mechanism among three flavor type of neutrinos. Such flavor-mixing mechanism, as shown in Eq. (16), could result in apparent oscillation behavior even if $m_0$ is exactly zero or extremely small.

## 3. Conclusions

In this work, we present a unified model that addresses three key questions in the Standard Model: why do leptons have three generations, why do neutrinos have vanishing masses, and why do neutrinos exhibit flavor mixing and mass oscillations? Our model utilizes geometric algebra operators in 4D spacetime and constructs relativistic energy and momentum operators for each generation of leptons using octonion operators. We begin by addressing the simple cases of a massless fermion and a point-like spin-1/2 particle with a rest mass in the first two sections. In the following sections, we expand our analysis beyond Dirac's theory for a point-like particle using the octonion operator approach. The conventional assumption that leptons and quarks are point-like particles without physical volumes is too simplistic to be logically consistent. This concept leads to a divergence in self-energy and does not explain why three generations of leptons exist or why their masses follow a simple Koide mass-ratio formula.

We have developed generalized wave equations for leptons that incorporate three additional momentum operators to describe their internal structural dynamics relative to the particle's center of mass reference frame. The conventional octonion form of the Zorn vector-matrix scheme only considers one k generation of a lepton, but in our unified modeling analysis, we have generalized this scheme to encompass all three generations of massive leptons and their neutrinos. We have also analyzed the relativistic mass-energy relation proposed by Einstein to explain why the rest mass energy of neutrinos is vanishingly small. We also show that the mass for neutrinos is not an eigenvalue and explain why flavor mixing and mass oscillation behavior occur among the three types of neutrinos. We explain, using Equations (13A-13C), why there exist exactly three generations of leptons, and we present in Eq. (15) the generalized Dirac equation using octonion operators that include three extra internal degrees of freedom. Moreover, we present, according to Equations (14A-14C), why there also exist three generations of neutrinos, and why they have vanishingly small masses. We present in Eq. (16) the generalized Dirac equation for neutrino, showing the source for flavor mixing and mass oscillations. In this work, we have provided physical explanations me unsolved issues about leptons that are beyond description of the Standard Model. Without the ad hoc assumptions in the Standard Model, our unified model provides a better and more natural understanding of some unclear or unsolved issues in the Standard Model. We are presently working on extending this geometric algebra operator approach to bosonic force carriers in the Standard Model, such as gluons and quarks, and will publish the results elsewhere.

## Acknowledgment

This research was funded by [Scientific Research Foundation for High-level Talents of Anhui University of Science and Technology] grant number [2022yjrc67].